\documentclass[conference]{IEEEtran}

\usepackage{amsmath,amsthm,amssymb,color,url}
\usepackage{graphicx}

\definecolor{grey}{gray}{.5}

\theoremstyle{plain}
\newtheorem{theorem}{Theorem}

\newtheorem{lemma}[theorem]{Lemma}

\theoremstyle{definition}

\newtheorem*{definition}{Definition}

\newcommand{\ii}{\mathrm{i}}

\renewcommand{\vec}{\mathbf}

\newcommand{\RR}{\mathbb{R}}
\newcommand{\bR}{\mathbb{R}}

\DeclareMathOperator{\EE}{\mathbb{E}}
\DeclareMathOperator{\Ex}{\mathbb{E}}
\DeclareMathOperator{\PP}{\mathbb{P}}
\DeclareMathOperator{\Prob}{\mathbb{P}}

\DeclareMathOperator{\Var}{Var}
\DeclareMathOperator{\Cov}{\text{Cov}}

\newcommand{\link}{\!\!\to\!\!}

\newcommand{\given}{\mid}

\newcommand{\cdec}{c_{\text{att}}}
\newcommand{\csep}{c_{\text{sep}}}

\newcommand{\SNR}{\mathtt{SNR}}
\newcommand{\INR}{\mathtt{INR}}

\newcommand{\snr}{\mathtt{snr}}

\newcommand{\Csum}{C_\Sigma}

\begin{document}

\title{Asymptotic Sum-Capacity of Random Gaussian Interference Networks Using Interference Alignment}

\author{\IEEEauthorblockN{Matthew Aldridge \qquad Oliver Johnson}
\IEEEauthorblockA{Department of Mathematics\\
University of Bristol, UK\\
$\{$m.aldridge, o.johnson$\}$@bristol.ac.uk}
\and
\IEEEauthorblockN{Robert Piechocki}
\IEEEauthorblockA{Centre for Communications Research\\
University of Bristol, UK\\
r.j.piechocki@bristol.ac.uk}}

\maketitle

\begin{abstract}
We consider a dense $n$-user Gaussian interference network formed by paired transmitters and receivers placed independently at random in Euclidean space. Under natural conditions on the node position distributions and signal attenuation, we prove convergence in probability of the average per-user capacity $\Csum/n$ to $\frac12 \EE \log(1 + 2 \SNR)$.

The achievability result follows directly from results based on an interference alignment scheme presented in recent work of Nazer et al. Our main contribution comes through the converse result, motivated by ideas of `bottleneck links' developed in recent work of Jafar. An information theoretic argument gives a capacity bound on such bottleneck links, and probabilistic counting arguments show there are sufficiently many such links to tightly bound the sum-capacity of the whole network.
\end{abstract}



\section{Introduction}

Recently, progress has been made on many-user approximations to the sum-capacity
$\Csum$ of random Gaussian interference networks.

In particular, in a 2009 paper, Jafar \cite{jafar} proved a result on the
asymptotic sum-capacity of a particular random Gaussian interference network:

\begin{theorem}[\cite{jafar}, Theorem 5]
  Suppose direct $\SNR$s are fixed and identical, so $\SNR_i = \snr$ for all $i$,
  and suppose that all $\INR$s are IID random and supported on some neighbourhood
  of $\snr$.  Then the average per-user capacity $\Csum/n$ tends in probability
  to $\frac12 \log(1+2\snr)$ as $n\to\infty$.
\end{theorem}

(Here and elsewhere, we use $\Csum$ to denote the sum-capacity of the network,
and interpret $\Csum/n$ as the average per-user capacity.)

A subsequent result by the current authors \cite{johnson} concerned a more
physically realistic model:

\begin{theorem}[\cite{johnson}, Theorem 1.5]
  Suppose receivers and transmitters are placed IID uniformly at random on the unit square $[0,1]^2$,
  and suppose that signal power attenuates like a polynomial in $1/\mathrm{distance}$.
  Then the average per-user capacity $\Csum/n$ tends in probability
  to $\frac12 \EE \log(1+2\SNR)$ as $n\to\infty$.
\end{theorem}


In this paper, we prove a similar -- but more general -- result to Theorem 2, with a neater proof, using ideas
from Jafar's proof of Theorem 1.  We assume transmitters and receivers are situated independently at random
in space (not necessarily uniformly), and that the power of signals depends in a natural way on the distance they travel.

Specifically our result is the following (full definitions of non-italicised technical terms are in Section II):

\begin{theorem} \label{thm:main}
  Consider a Gaussian interference network formed by $n$ pairs of 
  nodes placed in an \emph{spatially-separated IID network} with \emph{power law attenuation}.
  Then the average per-user capacity $\Csum/n$ converges in probability to 
  $\frac12 \EE \log(1 + 2 \SNR)$, in that for all $\epsilon > 0$
    \[  \PP \left( \left| \frac{\Csum}{n} - \frac12 \EE \log(1 + 2 \SNR) \right| > \epsilon \right)
          \to 0 \quad \text{as $n \to \infty$.} \]
\end{theorem}

The direct part of the proof uses interference alignment.   Interference alignment is a new
way of dealing with interference in networks, particularly when that interference is of a similar
strength to the desired signal.  Interference alignment allows communication at faster rates
than traditional resource division strategies such as time-division or frequency-divibysion
multiple-access.  Two early papers on interference alignment are those by Maddah-Ali, Motahari
and Khandani \cite{motahari} and Cadambe and Jafar \cite{cadambejafar}.

Specifically, we take advantage of so-called ergodic interference alignment, developed by
Nazer, Gastpar, Jafar and Vishwanath \cite{nazer}.

The converse part of the proof uses the idea of `bottleneck links' developed by Jafar
\cite{jafar}. An information theoretic argument gives a capacity bound on such bottleneck
links, and probabilistic counting arguments show there are sufficiently many such links
to tightly bound the sum-capacity of the whole network.

A different approach towards finding the capacity of large communications
networks is given by the deterministic approach of Avestimehr, Diggavi and Tse \cite{deterministic}. This paper
shows how capacities can be calculated up to a gap determined by the
number of users $n$, across all values of $\snr$. However, we identify a sharp
limit as the number of users $n$ tends to infinity.

A wider literature review is available in our previous paper \cite{johnson}.

The plan of this paper is as follows: In Section II we define our network model. In
Section III we prove the direct part of Theorem 3. Our main contribution comes in Section
IV where we use new ideas to prove the converse part of Theorem 3.  We conclude in Section V.

\section{Model}

\subsection{Node position model}

We believe that our techniques should work in a variety of models for the node positions. We outline
one very natural
scenario here.

These ideas were introduced in our earlier paper \cite{johnson}, but were not
fully exploited, due to that paper's concentration on the uniform case.

\begin{definition} \label{def:nodeplace}
Consider two probability distributions $\PP_T$ 
and $\PP_R$ defined on $d$-dimensional space $\RR^d$. Given an integer $n$, we sample the 
$n$ transmitter node positions
$\vec T_1, \ldots, \vec T_n$ independently from the distribution $\PP_T$. Similarly, we sample
the $n$ receiver node $\vec R_1, \ldots, \vec R_n$ positions independently from distribution $\PP_R$. We
refer to such a model of node placement as an `IID network'.
\end{definition}

Equivalently, we could state that transmitter and receiver positions are distributed according to two independent
(non-homogeneous) Poisson processes, conditioned such that there are $n$  points 
of each type.
 We pair the transmitter and receiver nodes up so that transmitter $i$ at $\vec T_i$ wishes to communicate 
with receiver $i$ at $\vec R_i$ for each $i$.
We make the following definition:

\begin{definition} \label{def:spatsep}
Let $\vec T \sim \PP_T$ and $\vec R \sim \PP_R$ be placed independently in $\RR^d$.
We say the IID network is spatially separated if there exists constants $\beta$
and $\csep$ such that for all $\rho$
  \[ \PP( \| \vec T - \vec R \| \leq \rho) \leq \csep \rho^{\beta} . \]
\end{definition}

In particular, it can be shown \cite[proof of Lemma 2.2i]{johnson} that the standard dense network
is spatially separated.
(The $d$-dimensional standard dense network is defined by $\PP_T$ and $\PP_R$ being
independent uniform measures on $[0,1]^d$.)
The standard dense network has been the subject of much research (see for example
the review paper of Xue and Kumar \cite{xuekumar} and references therein).  However, we emphasise
that our result holds for a wider range of models.

\subsection{Transmission model}

Our results are in the context of so-called `line of sight'
communication models, without multipath interference. That is,
we consider a model where signal strengths attenuate
deterministically with distance according to
some function $a$. 

The definitions in this section are adapted from our previous paper \cite{johnson}.

\begin{definition}
\label{def:transprot}
Fix transmitter node positions $\{ \vec T_1, \ldots, \vec T_n \} 
\in \RR^d$ and receiver node positions
$\{\vec R_1, \ldots, \vec R_n \} \in \RR^d$, and consider Euclidean distance
$\| \quad \|$ and an attenuation function $a$.
We define $\SNR_i = a(\|\vec T_i-\vec R_i\|)$, and for all pairs with $i \neq j$, define 
$\INR_{ij} = a(\|\vec T_i-\vec R_j\|)$.
\end{definition}

We consider the $n$-user Gaussian interference network defined so that
transmitter $i$ sends a message encoded as a string of $T$ complex
numbers $\vec{x}_i = (x_i[1], \ldots, x_i[T])$ to receiver $i$, under a power constraint
$\frac{1}{T} \sum_{t=1}^T |x_i[t]|^2
 \leq 1$ for each $i$.
 
Our result requires that the fading random variables be circularly symmetric.
For definiteness, we hold the modulus constant and choose the argument uniformly
at random.  (We discuss Rayleigh fading in Section V.)  So the $t$th symbol received at receiver $j$ is given as
\begin{multline} \label{eq:transmod} Y_j[t] =  \exp(\ii \Theta_{jj}[t]) 
\sqrt{\SNR_{j}} x_j[t] \\
+ \sum_{i\neq j} \exp(\ii \Theta_{ij}[t]) 
\sqrt{\INR_{ij}} x_i[t]  + Z_j[t],
\end{multline}
where noise terms $Z_j[t]$ are independent standard complex Gaussian random variables,
and the phases $\Theta_{ij}[t]$ are
independent $U[0,2\pi]$ random variables independent of all other terms.
The $\INR_{ij}$ and $\SNR_i$ remain fixed over time, since the node positions themselves are fixed,
but the phases are fast-fading, in that they are renewed for each $t$.

\begin{definition} \label{def:decay}
We say an attenuation function $a$ has power law attenuation if there exist constants
$\alpha$ and $\cdec$ such that for all $\rho$
  $ a(\rho)  \leq \cdec \rho^{-\beta} $
\end{definition}
 
(Tse and Viswanath \cite[Section 2.1]{tsevis} discuss a variety of models under which 
power law attenuation is an appropriate model for different exponents $\alpha$.)

For brevity, we write $S_{ij}$ for the random variables $\frac12 \log(1 + 2 \INR_{ij})$ (when $i \neq j$),
and $S_{ii}$ for $\frac12 \log(1 + 2 \SNR_{i})$
which  are functions of the distance
between the transmitters and receivers.
 In particular, since the nodes are positioned
independently, under this model
the random variables $S_{ij}$ are identically distributed, and $S_{ij}$ and
$S_{kl}$ are IID when $\{i,j\}$ and $\{k,l\}$ are disjoint. 

We will also write
$E = \EE S_{ii} = \frac12 \EE \log(1 + 2 \SNR)$,
noting that this is independent of $i$.  (It is also true that
$E = \EE S_{ij}$ for all $i$ and $j$.)


\section{Proof: direct part}

We can now prove our main theorem, Theorem 3, by breaking the probability into two terms which we deal
with separately. So
\begin{multline} \label{eq:unionbd}
 \PP \left( \left| \frac{\Csum}{n} - E \right| > \epsilon \right) \\
 = \PP \left( \frac{\Csum}{n} - E  < -\epsilon \right) +
\PP \left( \frac{\Csum}{n} - E  > \epsilon \right).\end{multline}

Bounding the first term of (\ref{eq:unionbd}) corresponds to the direct part of the proof.
Bounding the second term of (\ref{eq:unionbd}) corresponds to the converse part, and
represents our major contribution.

We prove the direct part as previously \cite{johnson}.

\begin{IEEEproof}
The first term of (\ref{eq:unionbd}) 
can be bounded relatively simply, using an achievability
argument based on an interference alignment scheme presented 
by Nazer, Gastpar, Jafar and Vishwanath \cite{nazer}. 
Their theorem \cite[Theorem 3]{nazer} implies that the rates $R_i =
1/2 \log(1 + 2 \SNR_i) = S_{ii}$ are simultaneously achievable. This implies that
$\Csum \geq \sum_{i=1}^n R_i = \sum_{i=1}^n S_{ii}$. This allows us to bound the first term in
(\ref{eq:unionbd}) as 
  \[ \PP \left( \frac{\Csum}{n} - E  < -\epsilon \right)
       \leq \PP \left( \frac{\sum_{i=1}^n S_{ii}}{n}   < E-\epsilon \right).
\label{eq:lln}
\]
But $E = \EE S_{ii}$, so this probability tends to $0$ by the weak law of large numbers.
\end{IEEEproof}

\section{Proof: converse part}

We now need to show that the second term of (4) tends to $0$ too.
Specifically, we must prove the following: for all $\epsilon > 0$
  \begin{equation} \label{converse}
    \Prob \left( \frac{C_\Sigma}{n} \geq E + \epsilon \right)
                                                              \to 0 
  \end{equation}
as $n \to \infty$.

The proof of the converse part is the major new part of this paper.
First, bottleneck links are introduced, and we prove a tight information-theoretic
bound on the capacity of such links.  Second, a probabilistic counting argument
ensures there are (with high probability) sufficiently many bottleneck links to bound
the sum-capacity of the entire network.

\subsection{Bottleneck links}

The important concept is that of the bottleneck link, an idea first used
by Jafar \cite{jafar} and later adapted \cite{johnson} in the following form:

\begin{definition}
  We say the link $i\link j$, $i\neq j$, is a \emph{$\epsilon$-bottleneck
  link}, if the the following three conditions hold:
    \begin{description}
      \item[B1:] $S_{ii} \leq E + \epsilon/2$,
      \item[B2:] $S_{ij} \leq E + \epsilon/2$,
      \item[B3:] $S_{jj} \leq S_{ji}$.
    \end{description}
  
  We let $B_{ij}$ be the indicator function that the crosslink $i\link j$ is a $\epsilon$-bottleneck
  link.  We also define the \emph{bottleneck probability} $\beta :=
  \Ex B_{ij}$ to be the probability that a given link is an $\epsilon$-bottleneck which is
  independent of $i$ and $j$ for an IID network.  (We suppress the $\epsilon$ dependence for simplicity.)
\end{definition}

The crucial point about bottleneck links is the constraints they place
on achievable rates in a network.

\begin{lemma} \label{links}
  Consider a crosslink $i\link j$ in a $n$-user Gaussian interference
  network.  If $i\link j$ is a $\epsilon$-bottleneck link, then the
  sum of their achievable transmission rates is bounded by
  $r_i + r_j \leq 2E + \epsilon$.
\end{lemma}

\begin{IEEEproof}
  First, note that we make things no worse by considering the two-user
  subnetwork:
    \begin{align*}
      Y_i &= \exp (\ii\Theta_{ii}) \sqrt{\SNR_i}X_i
               + \exp (\ii\Theta_{ji}) \sqrt{\INR_{ji}}X_j + Z_i \\
      Y_j &= \exp (\ii\Theta_{ij}) \sqrt{\INR_{ij}}X_i
               + \exp (\ii\Theta_{jj}) \sqrt{\SNR_j}X_j + Z_j
    \end{align*}
  where receiver $i$ needs to determine signal $X_i$, and receiver $j$
  signal $X_j$.  (The time index is ommited for clarity.)
  
  From bottleneck conditions B1 and B2 we have
    \[ 1 + 2\SNR_i \leq \exp (2E + \epsilon) , \quad
       1 + 2\INR_{ji} \leq \exp (2E + \epsilon) .     \]
  Summing and taking logs gives
    \begin{equation} \label{logs}
      \log (1+\SNR_i+\INR_{ji}) \leq 2E + \epsilon .
    \end{equation}
    
  We combine this with the argument given by Jafar \cite{jafar}.
  Let $r_i$ and $r_j$ be jointly achievable rates for the subnetwork.
  In particular, receiver $i$ can determine signal $X_i$ with an arbitrarily
  low probability of error.  

  We certainly do no worse if a genie presents signal $X_i$ to receiver
  $j$ -- so assume $j$ can indeed recover $X_i$.  But condition B3
  ensures that it is easier for receiver $i$ to determine $X_j$ than
  it is for receiver $j$ (since the weighting is larger in the first
  case).  So since receiver $j$ can recover $X_j$ (as $r_j$ is achievable),
  receiver $i$ can recover $X_j$ also.
  
  Because receiver $i$ can determine both $X_i$ and $X_j$, these
  two signals must have been transmitted at a sum-rate no higher
  than the sum-capacity of the Gaussian multiple-access channel
  (see, for example, Cover and Thomas \cite[Section 14.3.6]{cover}).
  Hence,
    \[ r_i + r_j \leq \log(1+\SNR_i+\INR_{ji}) \leq 2E + \epsilon, \]
  where the second inequality comes from \eqref{logs}.
\end{IEEEproof}

\subsection{Three technical lemmas}

A few technical lemmas are required in order to prove 
\eqref{converse}.

First, we need to ensure that very high $\SNR$s are very rare (Lemma \ref{SNRs}).
Second, we need to show that bottleneck links will actually occur (Lemma \ref{beta}).
Last, we must show that the number of bottleneck links cannot vary
too much (Lemma \ref{Var}).

Under any network model where these three lemmas are true, our theorem will hold.
We emphasise that our model of IID networks with power law attenuation is one such
model; we believe the result holds more widely.

\begin{lemma} \label{SNRs}
  Consider a spatially-separated IID network, with power law attenuation.  Then for any $\eta > 0$,
    \[ \Prob \bigg(\max_{1\leq i\leq n} S_{ii} > n^{\eta/2}\bigg) = O(n^{-1})
                                     \quad \text{as $n \to \infty$.} \]
\end{lemma}

In fact, in our case the convergence to $0$ is considerably quicker than
$O(n^{-1})$, but this is sufficient.

It is worth noting that this fast decay in the tails of $S_{ii}$ ensures that the
expectation $E = \EE S_{ii}$ does indeed exist and is finite.

\begin{IEEEproof}
First, we have by the union bound
  \[ \Prob (\max S_{ii} > n^{\eta/2}) \leq n \Prob ( S_{11} > n^{\eta/2}) . \]
But by the definition of $S_{11}$
  \begin{align*}
    \Prob ( S_{11} > n^{\eta/2})
      &= \Prob \left( \SNR_{11} > \frac12 (2^{2n^{\eta/2}}-1) \right)  \\
      &= \Prob \left( a(\| \vec T_1 - \vec R_1 \|) > \frac12 (2^{2n^{\eta/2}}-1) \right)
  \end{align*}
and the proof follows by applying the definitions of $\SNR$, spatial separation and
power law attenuation.
\end{IEEEproof}

We will often condition off this event; that is, condition on the complementary event
$\{ \max S_{ii} \leq n^{\eta/2} \}$.  We use $\PP_n$, $\EE_n$ and $\Var_n$ to
denote such conditionality, and write $\beta_n=\EE_nB_{ij}$ for the
conditional bottleneck probability.

The next two lemmas concern showing that conditional probabilities are
nonzero.  However,  we have for any event $A$,
	\begin{align*}
	  \Prob(A) &= \Prob(A \mid \max S_{ii} \leq n^{\eta/2}) \Prob(\max S_{ii} \leq n^{\eta/2}) \\
                  & \qquad\qquad {}+ \Prob(A \mid \max S_{ii} > n^{\eta/2}) \Prob(\max S_{ii} > n^{\eta/2}) .
  \end{align*}
and hence by Lemma \ref{SNRs} we have the bounds
  \begin{align*}
    \Prob(A) &\leq \Prob(A \mid \max S_{ii} \leq n^{\eta/2}) + \Prob(\max S_{ii} > n^{\eta/2}) \\
             &= \Prob_n(A) + O(n^{-1})
  \end{align*}
and 
	\begin{align*}
	  \Prob(A) &\geq\Prob(A \mid \max S_{ii} \leq n^{\eta/2}) \Prob(\max S_{ii} \leq n^{\eta/2}) \\
             &= \Prob_n(A) \big(1-O(n^{-1})\big),
  \end{align*}
and so $\Prob(A) = \Prob_n(A) + O(n^{-1})$.  This will be useful in the next two proofs.

\begin{lemma} \label{beta}
  Consider a spatially-separated IID network, with power law attenuation.  Then the conditional
  bottleneck probability $\beta_n$ is bounded away from $0$ for
  all $n$ sufficiently large.
\end{lemma}

\begin{IEEEproof}
  First note that by the comment above, we need only show that the
  unconditional bottleneck probability $\beta$ is nonzero.

  Second, note that by the exchangeability of $\vec R_i$ and $\vec R_j$,
  we have
    $ \Prob(\text{B1 and B2 and B3}) \geq \frac12 \Prob(\text{B1 and B2}) $.
  It is left to show that $\Prob(\text{B1 and B2})$ is non-zero.
  
  Note that B1 requires $S_{ii}$ to be less than its expectation plus $\epsilon$.
  So $\vec R_i$ must be situated such that this has nonzero probability.  So $\vec T_i$
  has a nonzero probability of being positioned such that B1 occurs.  But
  $\vec T_i$ and $\vec T_j$ are also exchangeable, so we are done.
\end{IEEEproof}

\begin{lemma} \label{Var}
  Consider a spatially-separated IID network, with power law attenuation.
  Then, conditional on $\{\max_i S_{ii} < n^{\eta/2} \}$,
  \[ \Var_n (\# \text{\emph{ bottleneck links}})
       = \Var_n \left( \sum_{i\neq j} B_{ij} \right)
       = O(n^3) ,                                  \]
  where the sum is over all crosslink pairs $(i,j)$, $i\neq j$.
\end{lemma}

In general, one might assume that $\Var_n (\# \text{ bottleneck links})$
would be proportional to the total number of links, and thus be $O(n^4)$.
However, because of the independences in the IID network, the variance is in fact
much lower.

\begin{IEEEproof}
  First consider the unconditional version.  We have
    \[ \Var \left( \sum_{i\neq j} B_{ij} \right)
         = \sum_{i\neq j} \sum_{k\neq l} \Cov(B_{ij}, B_{kl}) . \]
  The important observation is that for $i,j,k,l$ all distinct,
  $B_{ij}$ and $B_{kl}$ are independent giving $\Cov(B_{ij}, B_{kl}) = 0$.
  (This is because they depend only on the position of distinct and
  independently-positioned nodes.) Hence there are only $O(n^3)$ non-zero
  terms in the sum, each of which is trivially bounded by
  $\frac12 (1-\frac12) = \frac14$.
  
  But by the comment above, if $\Cov(B_{ij}, B_{kl}) = 0$, then the
  conditional covariance is $\Cov_n(B_{ij}, B_{kl}) = O(n^{-1})$.
  Hence, 
    \[ \Var_n \left( \sum_{i\neq j} B_{ij} \right) \leq O(n^3)\frac14 + O(n^4)O(n^{-1}) = O(n^3), \]
  as desired.
\end{IEEEproof}

\subsection{Completing the proof of Theorem 3}

We are now in a position to prove \eqref{converse}, and hence prove
Theorem 3.

\begin{IEEEproof}
We need to show
  \[
    \forall\, \epsilon > 0 \  \forall\, \delta>0 \ \exists\, N
       \  \forall\, n\geq N \quad
         \Prob\left( \frac{C_\Sigma}{n} \geq E + \epsilon \right)
           \leq \delta.      
  \]
So choose $\epsilon>0$, $\delta>0$, fix $n \geq N$ (where $N$ will
be determined later), and pick a rate vector $\mathbf{r}
\in \bR_+^n$ with sum-rate
  \begin{equation} \label{sumrate}
    \frac{r_\Sigma}{n} > E + \epsilon ;
  \end{equation}
we need to show that $\Prob(\mathbf r\text{ is achievable}) \leq \delta$.
(Here, we are writing $r_\Sigma := \sum_{i=1}^n r_i$ for the sum-rate.)

We divide into two cases: when there is a very high $\SNR$, which is unlikely
to happen; and when there is not, in which case $\mathbf r$ is unlikely to be
achievable.  Formally,
  \begin{align}
    &\Prob(\mathbf r \text{ achievable}) \notag \\ 
       &\ = \Prob(\mathbf r\text{ achievable} \given \max S_{ii} \leq n^{\eta/2})
           \Prob(\max S_{ii} \leq n^{\eta/2}) \notag \\
         &\ \quad\ {}+ \Prob(\mathbf r\text{ achievable} \given \max S_{ii} > n^{\eta/2})
             \Prob(\max S_{ii} > n^{\eta/2}) \notag \\
       &\ \leq \Prob(\mathbf r\text{ achievable} \given \max S_{ii} \leq n^{\eta/2}) 
       				+ \Prob(\max S_{ii} > n^{\eta/2}) \notag \\
       &\ \leq \Prob_n (\mathbf r\text{ achievable} )
           + \frac{\delta}{2} , \label{bound}
  \end{align}
for $N$ sufficiently large, by Lemma \ref{SNRs}.  We need to bound the first term in 
\eqref{bound}.

First, note that our assumption on $\max_i S_{ii}$ means that if $r_i > 2n^{\eta/2}$,
than we break the single-user capacity bound, since we would have
  \begin{multline*}
     r_i > 2n^{\eta/2} \geq 2 \max_j S_{jj} \\
       \geq 2 S_{ii} = \log (1+2\SNR_i) > \log (1+\SNR_i)
  \end{multline*}
meaning $\mathbf r$ is not achievable, and we are done.  Thus we assume this does
not hold; that
  \begin{equation} \label{boundR}
    r_i \leq 2n^{\eta/2} \quad \text{for all $i$}.
  \end{equation}

(The rest of our argument closely follows Jafar \cite{jafar}.)

Now, if $\mathbf r$ is achievable, it must at least satisfy the constraints on the
$\epsilon$-bottleneck links $i\link j$ from Lemma \ref{links}, and hence also the sum of those constraints.  So
  \begin{align}
    \Prob_n(\mathbf r &\text{ achievable}) \notag \\
      &\quad \leq \Prob_n ( r_i + r_j \leq 2E + \epsilon
          \text{ on bottleneck links $i\link j$} ) \notag \\
      &\quad \leq \Prob_n \left( \sum_{i\neq j} B_{ij} (r_i + r_j)
          \leq \bigg( \sum_{i\neq j} B_{ij} \bigg) (2E + \epsilon) \right) \notag \\
      &\quad = \Prob_n (U \leq V) , \label{UV}
  \end{align}
where we have defined
  \begin{align*}
    U &:= \frac{1}{n(n-1)} \sum_{i\neq j} B_{ij} (r_i + r_j) , \\
    V &:= \frac{1}{n(n-1)} \bigg( \sum_{i\neq j} B_{ij} \bigg) (2E + \epsilon) .
  \end{align*}

The conditional expectations of $U$ and $V$ are
  \[ \Ex_n U = 2\beta_n \frac{r_\Sigma}{n}, \quad
     \Ex_n V = \beta_n (2E + \epsilon)
                 = 2\beta_n \left(E + \frac{\epsilon}{2} \right). \]
Note that since $\beta_n > 0$ by Lemma 6, we can rewrite
\eqref{sumrate} as
  $ \Ex_n U > \Ex_n V + \beta_n\epsilon $ ,
or equivalently,
  \[ \Ex_n U - \frac{\beta_n\epsilon}{2} > \Ex_n V + \frac{\beta_n\epsilon}{2} . \]
The proof is completed by formalising the following idea: since 
the expectations are ordered $\Ex_n U > \Ex_n V$, we can only rarely
have the opposite ordering $U < V$.  Hence the expression in \eqref{UV} is small.

Formally, by (the conditional version of) Chebyshev's inequality and the union bound, we have
  \begin{align}
    &\Prob_n ( U \leq V) \notag \\
      &\quad \leq \Prob_n \bigg( U \leq \Ex_n U - \frac{\beta_n\epsilon}{2}
                    \text{ or } V' \geq \Ex_n V + \frac{\beta_n\epsilon}{2} \bigg) \notag \\
      &\quad\leq \Prob_n \bigg( |U - \Ex_n U| \geq \frac{\beta_n\epsilon}{2} \bigg)
              + \Prob_n \bigg( |U - \Ex_n U| \geq \frac{\beta_n\epsilon}{2} \bigg) \notag \\
      &\quad\leq \bigg(\frac{2}{\beta_n\epsilon}\bigg)^2 \Var_n U + \bigg(\frac{2}{\beta_n\epsilon}\bigg)^2 \Var_n V 
                                                                                                     \notag \\
      &\quad = \frac{4}{\beta_n^2\epsilon^2} ( \Var_n U + \Var_n V ) . \label{UV2}
  \end{align}

Using Lemma \ref{Var} we can bound these variances as
  \begin{align*}
    \Var_n U &= \frac{1}{n^2(n-1)^2} \Var_n \left( \sum_{i\neq j} B_{ij} (r_i + r_j) \right) \\
                 &\leq \frac{1}{n^2(n-1)^2} O(n^3) 16 n^\eta = O(n^{-(1-\eta)}). \\
    \Var_n V &= \frac{1}{n^2(n-1)^2} O(n^3) (2E+\epsilon)^2 = O(n^{-1}) ,
  \end{align*}
where we used \eqref{boundR} to bound $\Var_n U$.  Choosing $\eta$ to be
less than $1$, we can ensure $N$ is sufficiently large that for all
$n \geq N$
  \[ \Var_n U + \Var_n V \leq \frac{\beta_n^2 \delta \epsilon^2}{8} . \]


\noindent This makes \eqref{UV2} $ \Prob_n (U \leq V) < \delta/2$. 
Together with \eqref{UV} and \eqref{bound}, this yields the result.
\end{IEEEproof}

\section{Conclusion}

In this paper we have defined IID interference networks with power law attenuation.  We have shown
that this setup fulfils necessary properties for the average per-user capacity
$\Csum/n$ to tend in probability to $\frac12 \EE \log(1 + 2 \SNR)$.  We have
also noted that this result is not unique to our setup.

\newpage

We briefly mention one more example.
Suppose Rayleigh fading is added to our model.  That is, now let
$\SNR_i := |H_{ii}|^2 a(\|\vec T_i-\vec R_i\|)$ and
$\INR_{ij} := |H_{ij}|^2 a(\|\vec T_i-\vec R_j\|)$, where the $H_{ij}$
are IID standard complex Gaussian random variables.
Because ergodic interference alignment still works with Rayleigh fading \cite{nazer},
the direct part of the theorem still holds.  But also, because the fading coefficients
are IID, the independence structure from the non-fading case remains, ensuring Lemmas
5--7 hold.  Hence, the theorem is still true.

Characterising all networks for which such a limit for average per-user capacity
exists is an open problem.

At the moment, Theorem 3 should perhaps be regarded as being of
theoretical interest. That is, our major contribution is to provide a sharp
upper bound on the performance of interference networks. However, the
lower bound relies on an ergodic interference alignment \cite{nazer} which, while
rigorously proved, may not be feasible to implement in practice for large
number of users. Examination of the proof of the effectivenes of ergodic
interference alignment \cite[Theorem 1]{nazer} shows that,
even for a model with alphabet size $q$, the channel needs to be used
$O( (q-1)^{K^2})$ times. Even for $K \approx 10$, this is a prohibitive
requirement.
However, recent work by the current authors \cite{delay} characterises the delay--rate tradeoff
for ergodic interference alignment.
Also, note that for $K=3$, El Ayach, Peters, and Heath \cite{ayach}
have shown that the interference alignment scheme of Cadambe and Jafar \cite{cadambejafar} can perform
close to the theoretical bounds.
\section*{Acknowledgments}

M. Aldridge and R.~Piechocki thank Toshiba Telecommunications Research
Laboratory and its directors for supporting this work. The authors thank
Justin Coon and Magnus Sandell of Toshiba for their advice and support with this
research.


\end{document}